\documentclass{article}

\usepackage{arxiv}

\usepackage[utf8]{inputenc} 
\usepackage[T1]{fontenc}    
\usepackage{hyperref}       
\usepackage{url}            
\usepackage{booktabs}       
\usepackage{amsfonts}       
\usepackage{nicefrac}       
\usepackage{microtype}      
\usepackage{lipsum}
\usepackage{graphicx}
\usepackage{tabularx}
\usepackage{float}          

\begin{document}
\title{Nanoscale Connectomics Annotation Standards Framework}
 
\author{
    Nicole K. Guittari\textsuperscript{1}, Miguel E. Wimbish\textsuperscript{1}, Patricia \textbf{K. Rivlin\textsuperscript{1}, Mark A. Hinton\textsuperscript{1},} \\
    \textbf{Jordan K. Matelsky\textsuperscript{1}, Victoria A. Rose\textsuperscript{1}, Jorge L. Rivera Jr.\textsuperscript{1}, Nicole E. Stock\textsuperscript{1},} \\
    \textbf{Brock A. Wester\textsuperscript{1}, Erik C. Johnson\textsuperscript{1}, William R. Gray-Roncal\textsuperscript{1}} \\
    \\
    \textsuperscript{1}Research and Exploratory Development Department, \\
    Johns Hopkins University Applied Physics Laboratory, Laurel, MD, USA  
}
\maketitle
\begin{center}
\textbf{BENCHMARK Working Group.} Abhishek Bhardwaj, Alyssa Wilson, Ben Dichter, Bill Katz, Brock Wester, Chris Broz, Daniel Xenes, Erik Johnson, Hannah Gooden, Jordan Matelsky, Jorge Rivera, Josh Morgan, Kabilar Gunalan, Karl Friedrichsen, Kedar Narayan, Kushal Bakshi, Liam McCoy, Mark Hinton, Miguel Wimbish, Ming Zhan, Nicole Guittari, Nicole Stock, Norman Rzepka, Nuno da Costa, Oliver Ruebel, Pat Rivlin, Paul Fahey, Sahil Loomba, Sandy Hider, Stephan Gerhard, Tim Fawcett, Will Gray-Roncal, Yaroslav Halchenko
\end{center}

\begin{center}
\end{center}

\begin{abstract}
The promise of large-scale, high-resolution datasets from Electron Microscopy (EM) and X-ray Microtomography (XRM) lies in their ability to reveal neural structures and synaptic connectivity, which is critical for understanding the brain. Effectively managing these complex and rapidly increasing datasets will enable new scientific insights, facilitate querying, and support secondary use across the neuroscience community. However, without effective neurodata standards that permit use of these data across multiple systems and workflows, these valuable and costly datasets risk being underutilized especially as they surpass petascale levels. These standards will promote data sharing through accessible interfaces, allow researchers to build on each other's work, and guide the development of tools and capabilities that are interoperable. Herein we outline a standards framework for creating and managing annotations originating and derived from high-resolution volumetric imaging and connectomic datasets, focusing on ensuring Findable, Accessible, Interoperable, and Reusable (FAIR) practices. The goal is to enhance collaborative efforts, boost the reliability of findings, and enable comparative analysis across growing datasets of different species and modalities. We have formed a global working group with academic and industry partners in the high-resolution volumetric data generation and analysis community, focused on identifying gaps in current EM and XRM data pipelines, and refining outlines and platforms for standardizing EM and XRM methods. This focus considers existing and past community approaches and includes examining neuronal entities, biological components, and associated metadata, while emphasizing adaptability and fostering collaboration. 
\end{abstract}


\section{Introduction}
High-resolution image data with associated annotations and segmentations is crucial for enhancing our understanding of sensorimotor processing, vision, and behavior. The connectomics community has been at the forefront of developing methods for processing high-resolution image volumes and reconstructing the morphology of neurons and mapping neural circuits at the level of synaptic connections. These connectomic studies have relied on volumetric data generated from high-resolution imaging technologies such as electron Microscopy (EM), X-ray microtomography (XRM), and x-ray holographic nanotomography (XNH) to provide detailed features of neuroanatomy that can be identified and categorized.  Moving forward, these data can be compared with those from other modalities and collections, and a growing number of datasets collected from studies examining development, post-injury response, and progression (and treatment) of disease \cite{behrens2012human}. A major goal, defined and resourced by the BRAIN Initiative, is to generate accurate, detailed maps and atlases of neural connectivity and cell types of multiple species. This requires the generation of large-scale data collections (and the advancement of technology necessary to scale up these collections) \cite{lichtman2014big, hawrylycz2023guide}.  An associated goal is to enable cross modality analysis of these data collections and atlases, to include comparisons of anatomical and structural pathways, cell types, and functional activity (for example \cite{briggman2011wiring}). This will result in a more comprehensive understanding of neuronal circuits and how information is processed and integrated within the brain \cite{macpherson2021natural}. Ultimately, a long-standing goal of this community is to map full mammalian connectomes \cite{fan2019brief}.  Underpinning all of these activities is the need to effectively compare data across modalities, spatial and anatomical locations, and definitions such as cell types.

To date, there are only two full connectomes, one derived from a \emph{Caenorhabditis elegans} published 1986 \cite{white1986structure, emmons2015beginning}, and one from a \emph{Drosophila melanogaster} published in 2024 \cite{dorkenwald2024neuronal}. The \emph{Caenorhabditis elegans}, consisting of 302 neurons, and the \emph{Drosophila melanogaster} consisting of 140,000 neurons underscore the challenges posed by the human brain's 86 billion neurons [13, 21]. This work aimed to understand how genes shape nervous system structure and how the nervous system generates behavior. Understanding brain connectivity may contribute to developing efficient algorithms, neural networks, and cognitive function support \cite{moghadam2019algorithmic}. This study initiated the field of connectomics and spurred subsequent research efforts (Figure 1). These include Larry Swanson’s progressive mapping of the rat brain and projects like Machine Intelligence from Cortical Networks (MICrONS). In addition, insights into neural circuits may advance our understanding of neural information processig \cite{zador2023catalyzing}. However, the complexity of neurons requires standardized nanoscale definitions for human and machine analysis of connectomes. Community-defined standards ensure data can be interpreted, indexed, and retrieved across platforms by scientists with varying neuroscience expertise. This standardization enhances reliability and interoperability, promoting broader adoption and facilitating future research and discoveries. 
\begin{figure}[ht] 
    \centering
    \includegraphics[width=\textwidth]{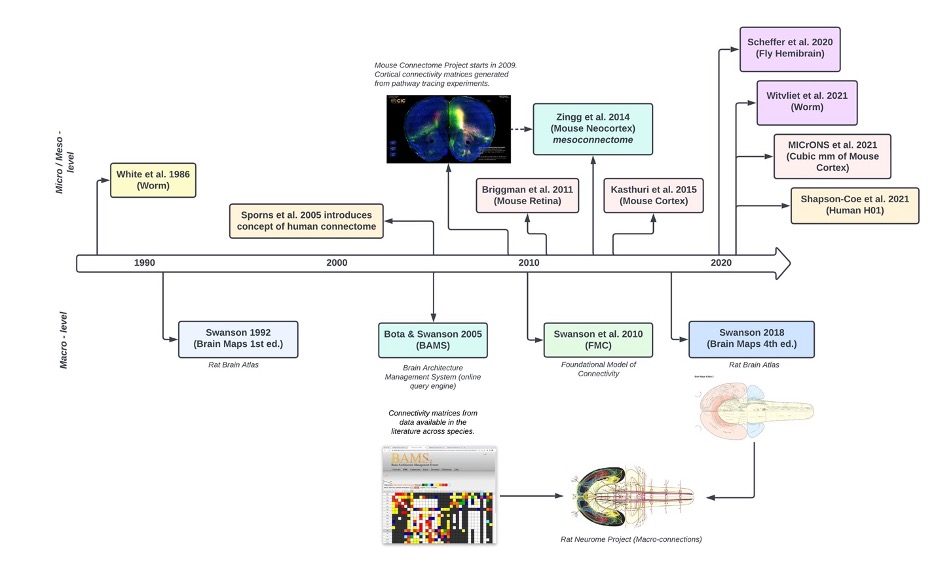} 
    \caption{Abbreviated timeline for resources and datasets used to map micro-, meso-, and macro-connections in the brain across species.} 
    \label{fig:figure1} 
\end{figure}

Neurodata standardization establishes a common language for researchers, enabling more efficient development and reducing costs. As data volumes grow, community-developed standards and platforms for studying and sharing neuronal entities will foster large-scale collaboration. Current neuroimaging standards in neuroscience, such as the Brain Imaging Data Structure (BIDS) \cite{gorgolewski2016brain, poldrack2024past} and the 3D Multi-scale Microscopic Image Standard (3D-MMS) \cite{ropelewski2022standard}, are integral to promoting global partnerships in the field. BIDS offers a systematic framework for data organization, enhancing data consistency and accessibility. These standards facilitate data reuse, collaborative research, and align with FAIR (Findable, Accessible, Interoperable, and Reusable) \cite{ropelewski2022standard} principles. 

Software platforms for connectomics like NeuVue, Neuroglancer, Connectome Annotation Versioning Engine (CAVE) \cite{dorkenwald2023cave}, NeuPrint \cite{plaza2022neu}, and NEURal Decomposition (NEURD) \cite{microns2021functional} complement standards development, broadening collaboration and analysis opportunities. Community-built platforms accommodate flexible annotations, enabling researchers to introduce new categories and establish uniform protocols and tools. By implementing these platforms, researchers can establish uniform brain definitions, protocols, and analysis tools. Such initiatives foster collaboration and standardization, driving progress in the field and enhancing educational approaches.

In fields like EM and XRM, having a single set of agreed-upon definitions is critical for ensuring interoperability across platforms. Multiple standards lead to issues like user inconvenience, confusion, and increased costs due to the need for different product versions and additional integration work. In neuroscience, where large-scale studies can cost millions, this inefficiency is particularly expensive. A lack of consensus can also slow down technological progress, as competing standards shift focus away from advancing the technology itself. 

To address the needs of the connectomics community, the Big-Data Electron-microscopy for Novel Community Hypotheses: Measuring and Retrieving Knowledge (BENCHMARK) project has formed a working group, drawn from academic laboratories, industry partners, and government agencies, to coordinate a common framework for the field of EM and XRM. The working group has developed standards proposals for 1) Image and Experimental Metadata and 2) Annotation Metadata. Here, we document the development of the BENCHMARK Annotation Metadata Standards for EM and XRM Connectomics v1.1. These standards define a set of characteristics and metadata structure that facilitates collaboration. We seek to evaluate a proposed set of biological structures, neuronal entities, annotation standards, and query-able interfaces for neurodata annotation exploration. Evolving definitions and software tools will be released and maintained at \url{https://github.com/aplbrain/benchmark-metadata} with implementations incorporated at \url{https://BossDB.org} to encourage adoption by the connectomics community.

\section{Development of Annotation Metadata Standards}
\label{sec:headings}
The brain’s complex structure often results in variability among researchers’ interpretation of neuroanatomical details, leading to diverse annotation outcomes even when analyzing the same datasets. This leads to discrepancies across datasets. This issue is compounded as the quantity and size of these datasets grows, with some exceeding petascale levels across hundreds of distinct datasets. Interpretation errors highlight the need for standardization to ensure accuracy and consistency in annotations. By applying specific definitions for neuronal structures, such as using flags to indicate the presence of synapses or dendrites and strings, we can establish a nuanced and flexible framework.  This approach under version control allows for refined secondary analysis and accommodates the unique classification requirements of neuronal entities. 

\subsection{Collaboration Efforts}
Identifying neural structures and functions creates a foundation for standardizing biological and computational entities. The BENCHMARK team organized a series of working group meetings with universities, laboratories, and neuroscientists, drawing from the international community, to begin to reach consensus on image, metadata, and annotation standardization for the field of EM and XRM connectomics. Continuous engagement from research facilities builds the platform for advancing the forefront of neuroscience. 
The BENCHMARK standards are being integrated with the BossDB archive to implement a prototype metadata API for the available projects. This initiative involves developing and balancing a set of hierarchical definitions alongside a user-friendly interface, bridging diverse fields and fostering a shared understanding for neurodata. Following our Experimental Metadata Standards, datasets are subdivided into projects, collections, experiments, and channels (in that order). Each entity has a corresponding title, description, grant number, publication information, and additional relevant statistics. The BENCHMARK team worked to manually validate information gathered from existing BossDB projects (\url{https://BossDB.org/projects}) eliminate data collection and formatting discrepancies. This was applied to experimental, imaging, and annotation metadata standard for the EM and XRM communities.

\subsubsection{Identification of Available Annotations}
Key variable discrepancies and their associated properties can be analyzed across publicly available datasets to identify repeated characteristics. Different key words are often used to annotate the same neuronal entity. Some datasets use the key word, “cell body”, while others use the key word, “soma” to refer to the cell body (Table 1).  Similarly, two datasets (Table 1) refer to “passant boutons” or “boutons”.  Discrepancies in key word vocabulary hinders comparative studies across datasets. This clearly demonstrates the need for standardization - if a statement is entered differently, a suggested query engine, even if flexible, could fail as a result. Variance in data classification and storage approaches, contribute to errors in neurodata exchange and incompatible query-able data. The proposed neuronal uniform standard (Figure 2), schema, and current manual efforts to sort available data allow users to filter based on key words and neuronal properties in a query-able metadata service. Without an agreed upon neuronal property specification, there will be continuous miscommunication, inconsistency, and misunderstandings in the field of connectomics. 
The proposed neuronal standards refer to cellular structures commonly annotated in neurons within EM connectomics datasets.  To identify these structures, we surveyed annotations associated with connectomics datasets stored on BossDB, as well as public datasets not stored on BossDB (Table 1).  Note that these standards are compatible with vertebrate and invertebrate datasets (e.g., mouse, drosophila).  While not shown in this version, this schema is also extensible to other cell types found in nervous tissue, such as glia and blood vessels.  As a wider range of subcellular structures are annotated in connectomics datasets, our schema can be integrated with schema that have a wider cell biological scope such as the Ontology of Subcellular Neuroanatomy \cite{larson2007formal}.  
In EM connectomics, several uniquely large datasets play a major role in how we identify optional neuronal entities. As connectomics becomes more accessible over time, there will be a continuous need to revise the standardization of data. To support current and future projects, a flexible yet simple annotation approach is essential. Mitigating confusion when entering data is difficult when scientists have different interpretations of the same structural entity (e.g., cellular compartments). Breaking down definitions to have both a type and value can help standardize to avoid this confusion. For instance, when asking a scientist to enter their corresponding data for an interneuron, one scientist may opt to enter the number of interneurons, and another may opt to enter the type of neuron. This calls for a separation in data entities, adding properties, such as “interneurons : Enum” and “interneuron-type: str” and enforces the use of explicit language when exchanging information. Though the use of strings and extensible enumerations allows for a flexible standard definition which can be reviewed and codified over time.

Eliminating the occurrence of the same entity in multiple classes also minimizes confusion. Entities should only be applied where and when appropriate, not in multiple classes as optional properties. For instance, “synapse” and “spine” are classes, but “synapse” includes spine-associated optional properties, such as spine ID (Figure 3). This situation may create duplicate pieces of information causing confusion and inconsistencies. Neurodata entities must therefore be organized to promote clarity and is evolving to support ongoing research (Figure 4). 

To enhance data accuracy, accessibility, and interoperability, metadata types are designed as specified enumerations (Enums) when possible. This approach allows for Enums to be assigned a set of named values and ensures that a specific name and numerical value can be associated. This facilitates interpretation of the value’s meaning and a robust method for representing a limited set of values.

Though previous research supports and will always support a degree of complex definitions, it halts widescale comparative analysis. The intricacy of neuronal properties and their corresponding metadata requires straightforward representations that should not be oversimplified to the cost of losing critical information, especially in the early developmental stages of connectomics. Researchers must find a balance between comprehensibility and capturing essential details. This concept is applied to diagram revisions in effort to create a cohesive understanding and groundwork that supports future discoveries.

\subsubsection{Elimination of Annotation Discrepancies}
Software interfaces for standardized annotation formats will improve data parsing and streamline neurodata management. Metadata is extensible through user definitions and software development is facilitated by providing a flexible standard in contrast to definitions. 

A common error present in both the creation of public metadata interfaces is the lack of methods to derive information. Ongoing efforts consist of manually gathering information from published literature and datasets, and these manual efforts contain thousands of lines of experimental properties. Though this query method allows for a legible interpretation of the data for projects, collections, etc., it is not effective for wide scale applications. Entering data by hand can introduce misspelling, errors in formatting, and slows collaboration between research facilities. Misspellings can be avoided by retrieving author information directly from publications or databases. Similarly, to types of separation found when entering data. Common errors arise from stylistic choices in data separation, such as the use of commas, bars, and spaces in query lists. From here, efforts have shifted away from manually sifting through text-based metadata, and towards adapting a persistent database supported by Mongo Express. This creates a platform for users to enter personalized data without having to push and pull a text-based metadata file (\url{https://metadata.bossdb.org/} ). 

In support of integrating compatible systems from BENCHMARK to BossDB and enhancing the usability and reliability of our data, the BossDB team has been developing and incorporating various open-source public tools (\url{https://bossDB.org/tools)}). These platforms, such as intern for data access \cite{matelsky2021integrated}, bossphorus as a simple volumetric datastore, and NeuVue for proofreading of machine segmentation, are instrumental in the data handling processes. Tools like pytri and substrate for visualization, and Jupyter notebooks for creating visualizations of co-registered cells, demonstrate steps towards sophisticated, user-friendly interfaces for neuroscientific data analysis. The Query Engine (\url{https://queries.bossdb.org/}) takes personalized query capabilities into account, allowing for the synthesis of connectomic dataset growth, while NeuPrint integration (\url{https://neuprint.bossdb.io/}) offers an ecosystem for analyzing connectome data. For network science applications, DotMotif (\url{https://github.com/aplbrain/dotmotif/}) provides a platform for subgraph isomorphisms on connectomes, and Grand is a scalable graph toolkit that enhances manipulability of data. Each of these tools reflects our commitment to address the need for more dynamic, collaborative, and error-resistant methods of data management, propelling neuroscience research into a new era of innovation. 
\begin{figure}[H] 
    \centering
\includegraphics{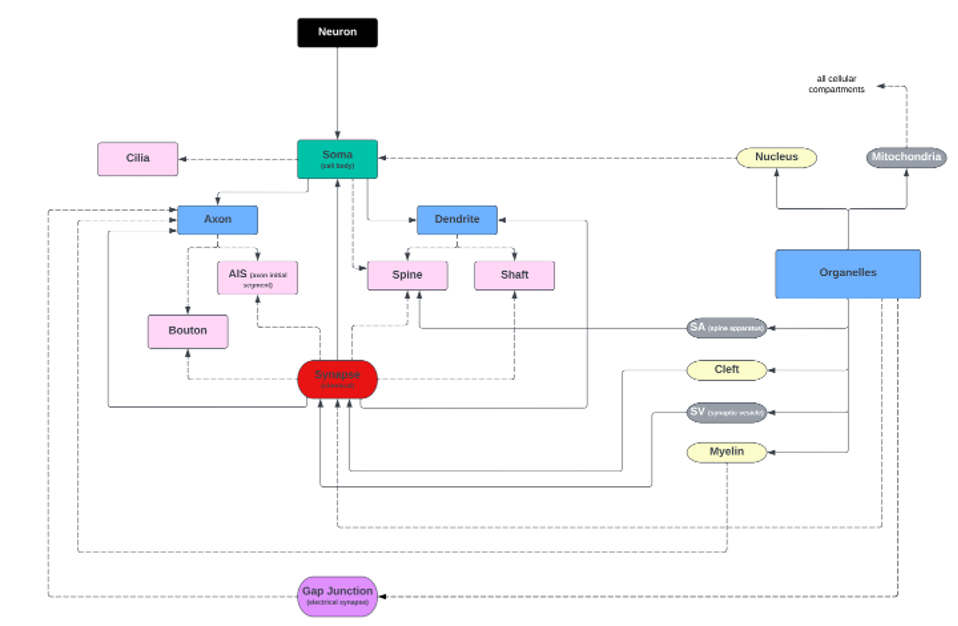}
     \caption{ Cohesive neuronal property connections with black indicating neuron type, green indicating soma, blue indicating cellular compartments and organelles, pink indicating axon-, dendrite-, and soma-specific properties, red indicating a synapse, purple indicating gap junction, yellow indicating organelle-specific classes, and grey indicating organelle entities commonly associated with synapses.} 
\end{figure}

\begin{figure}[H]
    \centering
    \includegraphics[width=0.6\textwidth]{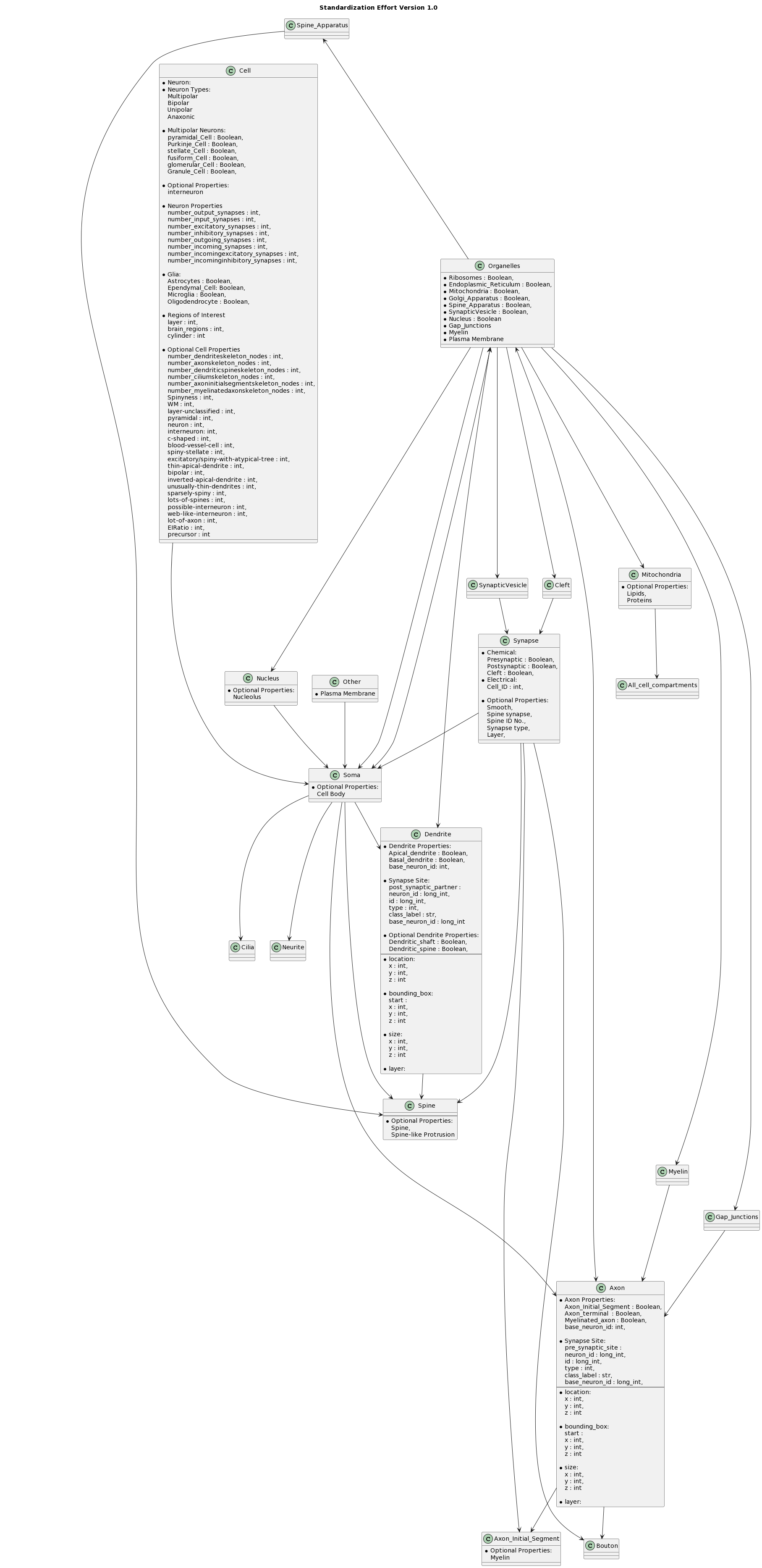} 
    \caption{Plant UML describing a proposed classification for neuronal properties, optional entities, and corresponding metadata type for Standardization Effort (Version 1.1).}
    \label{fig:figure3} 
\end{figure}

\begin{figure}[H]
    \centering
    \includegraphics[width=\textwidth]{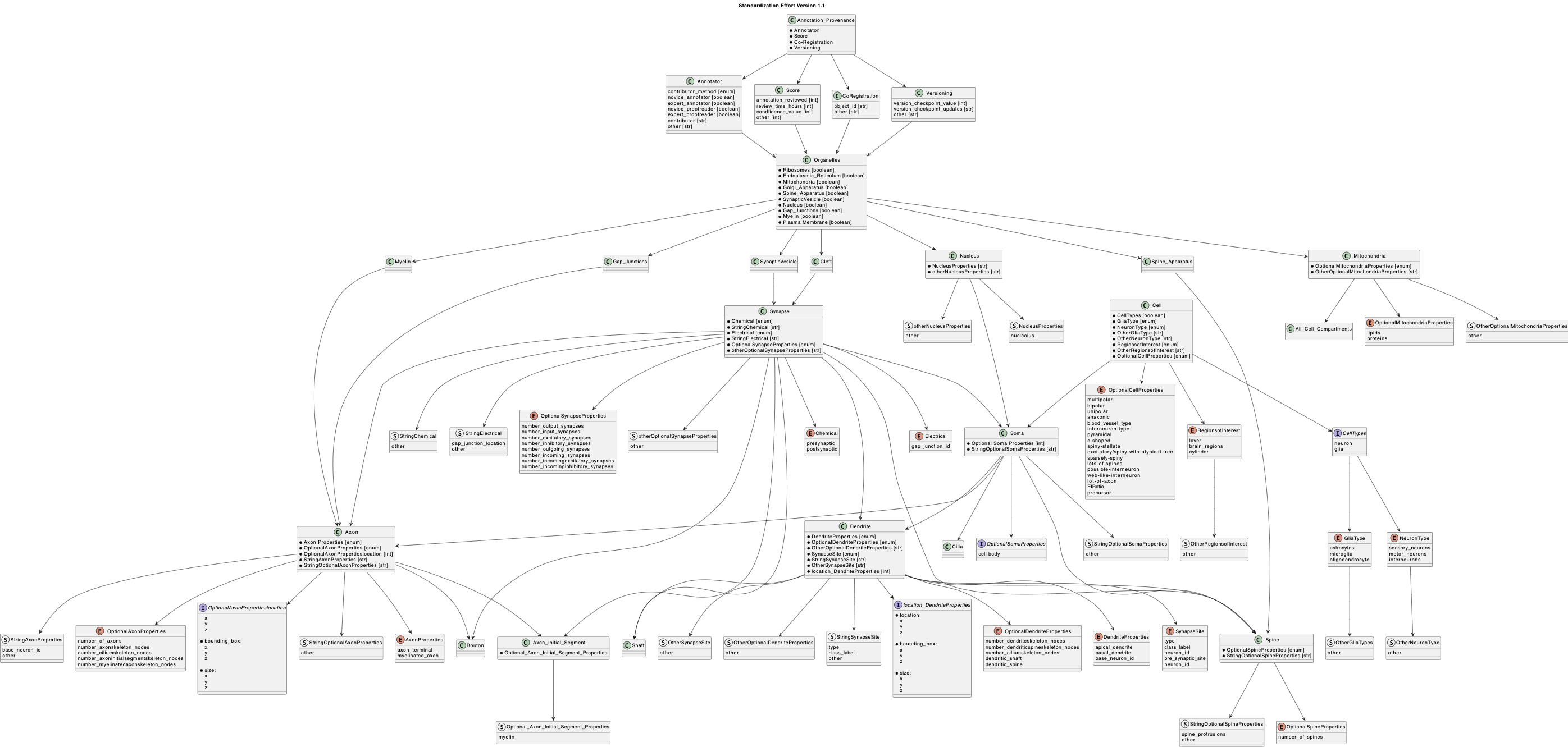} 
    \caption{Continuation of Plant UML in Figure 3, incorporating previous definitions and research to ensure broader compatibility. This hierarchical version control takes into account community engagement in efforts to democratize Standardization Effort (Version 1.1).}
    \label{fig:figure4} 
\end{figure}

\section{Results}

\subsection{Proposed Neurodata Structure}
To address the challenge of synonymous definitions within neuroanatomy, an adaptive framework (Figure 2) that proposes the architecture of neuronal components has been created. It serves as an instrumental guide in the categorization and labeling processes for metadata. The figure is color-coded: black lines map out neuron types, green signifies the soma, blue identifies cellular compartments and organelles, pink highlights axonal, dendritic, and somatic-specific properties, red shows synapses, purple denotes gap junctions, yellow points out organelle-specific classes, and grey marks entities commonly associated with synapses. This organizational schema establishes boundaries and classifications, aimed at aiding annotators in consistently identifying and recording neuronal structures, preventing the misclassification of identical entities. Figure 2 is not a static representation but an evolving tool, poised to be updated as our understanding of neuroanatomy expands, ensuring the standardization and clarity of data across the field under community-governed version control.

The designed anatomical structure is a baseline diagram of neuronal standardizations in our exchangeable effort to maintain consistent accuracy (Figure 2). Classifying anatomical entities differs amongst scientists, calling for a simplistic communal structure. Adopting a structure that organizes class diagrams can be used to depict relationships for each class to contain entities, data types, and their allowed values. Each class serves as a broad definition that can have multiple entities with corresponding data types, each representing a different piece of data associated with the neuronal class. Each data, integer, float, string, enumeration, Boolean or other, depends on the data being annotated. This fluid structure allows for classes to have multiple attributes of the same or different data types. For each neuroanatomical property in a dataset, such as, cell, organelle, axon, dendrite, etc., data owners can provide the user or community defined entities and optional properties where they are deemed fit (\url{https://github.com/aplbrain/BENCHMARK-Metadata/annotation-metadata/required-field-names.md}) for continued refinement. This effort aims to construct an explicit standard that has room for fluidity. 

Efforts have been pieced together based on research gathered from working groups and emerging publications.  This combination can be found on our GitHub page, where continuous open collaboration is publicly encouraged on the discussion page.  Collecting ongoing efforts serves as a valuable tool in mitigating confusion and creating version controls for academic partners to adopt, contribute, and work with. Continuous version control seeks to adopt and expand previously developed community frameworks. Our focus is on embracing and enhancing existing community frameworks like CAVE, NeuPrint \cite{plaza2022neu}, Neuroglancer, NEURD \cite{microns2021functional}, NeuVue \cite{xenes2022neuvue}, among others, while also remaining open to incorporating additional innovative systems. CAVE \cite{dorkenwald2023cave} provides a computational platform for reliable analysis of connectomes in large datasets with an advanced proofreading system for segmentation. This system ensures linkage of annotations to spatial coordinates, aligning them with corresponding segments, similar to how annotations correspond with metadata. NeuPrint is designed to handle the analysis of large-scale connectome data, offering tools and functionalities for this data-intensive field \cite{plaza2022neu}. Neuroglancer, utilizing WebGL, provides a platform for the visualization and viewing of 3D volumetric data. NEURD, built as a Python package, specializes in extracting 3D neuronal mesh data and facilitates a range of processes, such as automated proofreading, morphological examination, and studies in connectomics \cite{microns2021functional}. NeuVue is a software platform specifically developed for extensive proofreading and reconstruction of neural circuits in high-resolution electron microscopy connectomics datasets \cite{xenes2022neuvue}. It offers a sturdy web-based interface that enables collaborative viewing, annotating, and editing of segmentation and connectivity data by proofreaders. These tools collectively enhance the capabilities in connectome research and analysis. Choosing to implement community-built platforms accommodates schematized and adaptable annotations, granting researchers the ability to innovate and introduce new annotation categories while enhancing current systems (Figure 5).

Embracing community-developed platforms allows for structured and adaptable annotations, providing researchers with the flexibility to innovate and introduce novel annotation categories. This approach fosters collaboration and accommodates diverse research needs and methodologies. Pulling connectome data on a granular level and communicating the findings of research conducted will maximize time for conducting the studies as opposed to allocating extensive time on data procurement. These efforts are intended for users with varying levels of neuroscience background to familiarize themselves simplify the complexity of diagrams and tools that take into account neuronal details. 

\subsection{CAVE Integration and Schema Expansion}
In order to standardize and expand the annotation metadata framework for EM and XRM connectomics, we are actively working with CAVE schemas as a foundational template. Our primary objective is to address the gaps in current data annotation practices, ensuring that they are consistent, accurate, and scalable. By meticulously going through each class in the existing CAVE schema and adding annotations based on the BENCHMARK-defined classes, we are enhancing the capabilities of these schemas to better handle the growing complexity of large-scale datasets.

For example, by introducing detailed spatial and functional annotations, we are improving the precision and scope of data analyses. These updates make the schema robust and adaptable, enabling it to support new annotation types critical for accurately representing significant entities in large-scale EM datasets. Additionally, by extending and customizing these base schemas—such as BoundSpatialPoint for linking spatial locations to neuronal entities—we are ensuring that our proposed framework remains flexible and precise.

These efforts are part of a larger initiative to enhance data interoperability and collaboration within the neuroscience community. By making these enhancements, we aim to provide researchers with the tools they need to conduct more effective and insightful analyses, contributing to new discoveries and advancements in the field of connectomics.

\subsection{Community Feedback}
Working group attendees have been encouraged to contribute to these community efforts and voice critiques. With this, our goal is to foster a dynamic and growing community of academic partners that can validate and steer the development of beneficial milestones in the EM and XRM connectomic community. Outreach efforts consist of meeting individually with working group participants get an understanding for current annotation metadata approaches and needs, and a series of working group meetings for discussion of key topics. The virtual quarterly meetings consist of reviewing preliminary image and annotation standards gathered from the community and past datasets, breakouts to discuss use cases and requirements, breakouts to discuss metadata topics and data formats, gathering feedback on existing standards approaches, and developing release and maintenance plans. Through these efforts, we aim to not only advance the field but also to ensure that our approaches are continually refined and responsive to the evolving needs of the connectomics community.

\subsection{Working Meetings}
At the working group meetings, existing neuroanatomical annotations and proposed structures were reviewed and there was also discussion around the goals of standards development and implementation strategies. Proposed solutions included a flexible development process that is reusable across labs to combat the need to frequently rewrite code. Neuroscientists expressed the importance of creating a way to interact with existing data and the importance of interoperable formats between groups. Making efforts for open-source governance, the use of the Zarr file format, and overall accessibility were also high priority issues. Working group members suggested Git Annex as a method to address rapidly changing data. Consensus was reached that the developing standard must coincide with backward compatibility and stay compliant with varying data types to ensure accurate scalability and size.  

Further discussions centered around draft annotation and neuroanatomical standards, as well as existing software tools. Key discussion points included the limitations of connectomic data, incorporating established procedures for generating connectomic data, capturing the accuracy of connectomics data, and version control strategies for data. Approaches for version control must address both evolving annotations but also checkpoint versioning. Version control must also include all relevant files and their corresponding parts. All parts must indicate that the schema version is consistent throughout all representation. 

Further working group discussions centered around emerging technologies for connectomics data, for instance the use of AI to perform unsupervised learning at low resolutions (which has proven to be a flexible method to parse cell types). However, neurite classification and hierarchical definitions are critical for connectomics analysis and remain a challenge for AI approaches. Differing definitions between cell types were discussed and questioned if there is an example of one overall designation that could be used. The intricate details of neuroanatomical structures require a defined ontology of neuroscience entities in order to use an AI oriented system. This ontology must encompass the large array of cellular and metadata components. Overall, working group discussions demonstrated that a proposed neurodata standard was critical for the advancement of EM connectomics. 

\section{Discussion}
The BENCHMARK Annotation Metadata Standards represent a preliminary step towards standardizing EM connectomics, offering a comprehensive collection of definitions, exchangeable metadata interfaces, and a standardized, queryable neurodata library. These standards are continuously revised and adapted, ensuring they remain relevant and effective. This dynamic approach is essential for fostering growth and collaboration in the field of connectomics, as it encourages ongoing development and refinement of methodologies and technologies. Addressing these challenges will unify researchers to work towards a comprehensive understanding of the brain. 
Defining and classifying key properties differs between team members, and even more so between academic partners, slowing the creation of exchangeable metadata interfaces. Connectomics standards must contain version controls, hierarchal definitions, and open-source collaborative networks to exchange findings. Version control will organize data and monitor advancements, hierarchal definitions will eliminate misrepresentation, and collaborative networks will foster a motivated community. Even if the community determines common threads in standardization issues, implementation will differ. Platforms and discussion forums foster collaboration and reduce confusion in the community. Including all invested academic partners is a challenge, and language barriers, time zone restrictions, and schedules make it difficult to create inclusive global working groups. 
Creating a standard is in part based on specific project approaches and established preferences, adding to the complexity of organizing data. The nomenclature must be flexible, agreed upon, and explicit. To promote this, we seek to create open-source available data (\url{https://bossdb.org/projects}) and tools (\url{https://bossdb.org/tools}), encompassing a variety of species and intricate methods to access, process, network, and visualize connectomics. 
Neuroscience standards have the potential to be highly impactful. Advocating for broad use cases will ensure their adoption and evolution in diverse applications. This effort is part of an evolving neurodata annotation standard from BENCHMARK to keep pace with the rapidly progressing field of neuroscience. We anticipate that over time, this nomenclature will support emerging efforts such as the NIH BRAIN CONNECTS program, and hope that these discoveries can be applied to educational groups, grow research, spark motivation, and serve as a catalyst in our goal to create adaptable monumental neuroscience discoveries. 

\begin{table}[ht]
\centering
\caption{Annotations from Various Studies}
\begin{tabularx}{\textwidth}{|l|X|X|X|X|X|X|X|}
    \hline
    \textbf{Annotations} & \textbf{Kasthuri et al. 2015} & \textbf{Witvliet et al. 2020} & \textbf{Hemibrain} & \textbf{IARPA MICrONS Minnie} & \textbf{H01} & \textbf{Nguyen \& Thomas, et al. 2022} & \textbf{Prasad et al. 2020} \\ \hline
    Cell Body & Cell Body & Soma & Soma & Soma & Cell body & Soma & Cell body \\ \hline
    Region of Interest & Cylinder & Module & Brain Region & Layer & Layer & Region of Interest & Cortex, Striatum, Thalamus, and Zona Incerta \\ \hline
    Neuron & Neuron, Neuronal somata & Neuron & Neuron & Neuron & Neuron & Neuron & NA \\ \hline
    Bouton & Bouton & Passant boutons & NA & NA & NA & Bouton & NA \\ \hline
    Interneurons & Putative interneurons & Interneurons & NA & NA & NA & Molecular layer interneurons & NA \\ \hline
    Synapse & Synapse & Synapse & Synapse & Putative synapses & Synapse & Synapse & Synapse \\ \hline
    Synaptic Cleft & NA & NA & NA & Cleft & Clefts & Synaptic Clefts & Synaptic Cleft \\ \hline
    Spine Apparatus & Spine Apparatus & NA & NA & NA & NA & NA & NA \\ \hline
\end{tabularx}
\label{tab:annotations_table}
\end{table}

\section{Acknowledgments}
We would like to acknowledge and thank our BENCHMARK working group participants, listed above, for their invaluable contributions and expertise. Their contributions formed the core of the suggested standards, and we are honored to collaborate with these individuals. We would like to thank the BRAIN Informatics program archives, the BICCN consortium, the NWB standards group, the BIDS standards group, and the 3D-MMS group for developing and maintaining their valuable metadata standards. This work was supported in part by National Institutes of Health (NIH) grants R24MH114799, R24MH114785, and R01MH126684. The content is solely the responsibility of the authors and does not necessarily represent the official views of the National Institutes of Health.

\bibliographystyle{ieeetr}
\bibliography{template}

\end{document}